# A Formal Approach to Network/Distributed Systems Complex Testing

Andrey A. Shchurov, Radek Mařík, Vladimir A. Khlevnoy

*Department of Telecommunications Engineering, Faculty of Electrical Engineering,*

*Czech Technical University in Prague, The Czech Republic*

*Abstract*— **Deployment of network/distributed systems sets high requirements for procedures, tools and approaches for the complex testing of these systems. This work provides a survey of testing activities with regard to these systems based on standards and actual practices for both software-based and distribution (network) aspects. On the basis of this survey, we determine formal testing procedures/processes which cover these aspects, but which are not contrary to both aspects. The next step, based on the analysis of the implementation phase of System Development Life Cycle, determines a formal model for these processes.**

## I. INTRODUCTION

When talking about the testing of network/distributed systems, we face the great challenge of the dual nature of these systems. Today every computing and/or communication (networking) component is a computer with special operating software. So, it is necessary to pay respect to their software nature and we have to talk about *software-based systems* instead of simply *systems*. On the other hand, the distribution character of these systems forces us to consider their network nature. So, test applications should be very flexible to cover these systems appropriately and test multiple different aspects with a variety of requirements [1].

Historically, the software part is the domain of system and software engineers. Respectively, the network part is the domain of network engineers and partially for system engineers. As a consequence, system, software and network engineers have few common models or approaches and even their vocabulary is different [2]. The situation worsens when we talk about industrial control systems. Objectively, they have the same physical nature as each and every distributed system. But practically, the domain of industrial engineers was closed for system and/or network engineers for many years [3].

Fortunately, in the last few years there have been some positive changes:
- virtualization and cloud technologies inspire convergence vocabularies, models and approaches of system, software and network;
- the use of standard IT-technologies in industrial control systems responding to standardization and price reduction processes inspires the convergence of the industrial engineer community and other IT-specialists (system, software and network engineers).

But these changes are not yet completed.

Our main goal is the automated design and generation of testing procedures/specifications and plans for distributed systems as a necessary part of the project documentation. As the first step to this goal, we have to define the following formal structures:
- Testing processes. These processes have to be appropriate for both parts of systems – communication (network) subsystem and service/application (software-based) subsystem.
- Model of testing activities. This model must represent testing processes during the System Development Life Cycle (SDLC) and their interactions with other processes.

The rest of this paper is structured as follows. Section 2 introduces a foundation regarding testing activities, including network, software-based systems, performance and security testing approaches. Section 3 represents formal testing procedures that are appropriate for all aspects of network/distributed systems. In turn, Section 4 describes a formal model of test activities for network/distributed systems. Finally, conclusion remarks are given in Section 5.

## II. BACKGROUND

Based on the dual nature of distributed systems, we can consider two main aspects of activities for their testing: (1) network subsystem testing; and (2) software-based subsystem testing. Furthermore, we can consider two additional aspects due to their importance: (3) performance testing; and (4) security testing. In the following four subsections we deal with a brief overview of their related standards and practical methods.

### A. Network Testing

The current revisions of ISO/IEC 9646:1994 [4] and ITU-T Z.500 [5] standards define distributed systems as interconnection processing and, as a consequence, conformance testing methodology. Conformance testing is a powerful tool for networks protocol testing but it does not really cover network functionality [6].

Necessary complements to ISO/IEC 9646:1994 were done by Grabowski and Walter [7] [8] as a Test Methodology for Distributed Systems, which includes the following procedures:
- Conformance Testing;
- Interoperability Testing;
- Functional Testing;
- Performance Testing.

As a practical approach, Buchanan defined test objectives across the network life cycle [9]. This approach is based on practical experience (and remains relevant [10]), but it is not a systematic approach and Buchanan used the word *Art* in the title of his books. At present Cisco Network Life-Cycle Approach (PPDIOO) is a very effective practice. Based on PPDIOO, Ranjbar [11] and Sholomon and Kunath [10] define network testing procedures (as a Standard De-facto for Cisco Systems) for the different phases of the systems development process.

PPDIOO (like its predecessor [9]) is based on real practical experience, but not on standards or formal methods, and can be used as a case study for comparing formal and practical approaches.

B. *Software and System Testing*

Standards IEEE 1012:1998 [12], ISO/IEC/IEEE 12207:2008 [13], ISO/IEC/IEEE 15288:2008 [14] and ISO/IEC/IEEE 29119:2013 [15] define top-level test activities across the systems and software life cycle. The basic technical processes are:
- Software/System Acceptance Testing (Validation Process);
- Software/System Qualification Testing (Verification Process).

As a practical approach, we have application layer test objectives across the network life cycle [9] and other recent works on Internet-based/Network-based software testing [16]. Myers et al. [17] defines verification and validation processes for Internet (client-server) applications.

In contrast to the network testing, the recent works are based on both practical experience and on formal methods. It applies (at least theoretically) to the individual components and the whole system.

On the other hand, while traditional software testing focuses on code-level testing and evolves with Distributed and Web Service architectures, SOA application testing mostly introduces a testing of business logic [18]. But at the highest level, a testing of SOA implementations does not differ from the testing of traditional distributed (network) systems. Specifically, SOA testing has to address [19]:
- functionality (including performance);
- non-functional attributes (including interoperability);
- conformance.

C. *Performance Testing*

Performance characteristics are the most important source of information about network/distributed systems conditions. Based on this information, customers make their conclusion about these systems – whether they are ready for use or not. And nobody in the real world would tolerate high values of application response time. Moreover, performance testing is the key factor in the checking of system reliability [20]. As a consequence this area is well-defined.

The set of standards – RFC 2544 [21], RFC 2889 [22], RFC 3511 [23] and RFC 5180 [24] – defines the Benchmarking Methodology and, based on this methodology, network performance characteristics.

In turn, application performance requirements, include IP-based services, are defined by ITU-T Rec Y.1540 [25] and ITU-T Rec Y.1541 [26].

These standards cover all aspects of performance testing activities, including Quality of Services (QoS) aspects. But in practical terms, there are plenty of possible procedures defined by these standards and we need a systematic high-level approach based on practical experience.

Surprisingly, Buchanan [9] defines only two types of performance (or response time) testing activities: (1) application response time; and (2) throughput. Obviously this is not enough for the objective evaluation of the system behaviour - these two tests do not cover performance characteristics defined by standards.

An effective practical approach to performance test activities, which is the most relevant to network/distributed system testing, is defined by Jain [27]:
- Time-rate-resource testing:
  - Response time;
  - Throughput/Bandwidth;
  - Utilization.
- Reliability (error-free) testing.
- Availability (downtime) testing.

In spite of the year of publication, this approach is not contrary to the current revisions of standards.

D. *Security Testing*

Nowadays network/distributed systems have critical security requirements. Their failure may endanger human lives and the environment, do serious damage to major economic infrastructure, endanger personal privacy, undermine the viability of whole business sectors and facilitate crime [2]. As a consequence, the most difficult part of systems deployment is the question of assurance (whether the system will work) and verification. If assurance is difficult, verification is even more difficult – it is a question of how to convince customers (and, in extremis, a jury) that a system is indeed fit for its goals including security objectives.

The current revisions of ITU-T X. 1051 [28], ISO/IEC 27001:2013 [29], ISO/IEC 27002:2013 [30] and ISO/IEC 27005:2013 [31] standards define requirements for Information Security Management System (ICMS), code of practice for information security controls**,** risks of Information Security (IS) and risk management. The basic technical process is defined as Penetration Testing (or hacking).

As a practical approach, Yong and Aitel [32], Allen [33], Weidman [34] and others (there is a huge number of publications) define the methodology and the best practice for assessing network security by generating and executing possible attacks.

Similar to the software testing, the recent works are based on both practical experience and on formal methods. It applies (at least theoretically) to the individual components, the system itself and the system's environment.

III. FORMAL TESTING PROCESSES

Now we can determine test activities as a set of processes that meet our objectives, based on the following criteria: (1) these processes should be based on standards and actual practices, (2) they have to cover all aspects of network/distributed systems, (3) they have to be appropriate for all subsystems/components of these systems, and (4) they have to be simple enough for practical application.

Logically, as top-level processes we have:
- Validation testing processes;
- Verification testing processes.

In turn, the structure of verification testing processes can be represented as:
- Conformance testing processes;
- Interoperability testing processes;
- Functional testing processes;
- Performance testing processes;
- Security testing processes.

And the structure of the performance test activities is:
- Time-rate-resource testing processes;
- Reliability testing processes;
- Availability testing processes.

A recommended selection of test documentation needed to support these processes is defined by ISO/IEC/IEEE 29119-3:2013 ~\cite{ISO29119}.

## IV. FORMAL MODEL

Based on the software nature of network/distributed systems, the well-known Verification and Validation Model (V-Model) [35] can be used as a starting point. The V-Model layers reflect different viewpoints of testing at different layers of detail. And based on the distributed nature of these systems we can use OSI Reference Model [36] for the layers' description [37]:
- Physical architecture – Physical (L1) and Data Link (L2) Layers of OSI Reference Model - we cannot divide these layers in the case of commercial off-the-shelf (COTS) telecommunication/network equipment.
- Logical architecture – Network (L3) Layer of OSI Reference Model.
- Service architecture – Transport (L4), Session (L5), Presentation (L6) and Application (L7) Layers of OSI Reference Model – we cannot divide these layers in the case of COTS software.

And then the model must be adapted to our requirements – network/distributed systems testing. Just as every V-Model, it defines [35]:
- Top-Down design processes flow from End-User requirements to technical specification and installation & testing plan through Service, Logical and Physical architecture design layers.
- Bottom-Up installation and testing processes flow from the technical specification and installation & testing plan to system validation through Service, Logical and Physical topology verification layers.

However, one-way process flows can exist only in an ideal world. In the real world we have to foresee the following situations:

– *Design back-workflow.* If the challenges of a high-level architecture design solution cannot be resolved by a low-level design process (a very practical situation), we have to re-start the high-level design process and re-design this architecture solution. So, our adopted model should permit steps back from Physical architecture design layer up to End-User requirements. The last step back to End-User requirements implies problematic issues – it is very difficult and sometimes unrealistic to make some corrections in this document without penalties. As a consequence, the design process flow of the V-Model can be represented as the Waterfall model [38].

– *Horizontal relations from installation/testing processes to design process.* These relations are defined by the existence of problems which must be resolved – challenge requests in the small model. There are two main sources of these problems:
- Human factor:
  - Errors in the design solutions;
  - Ineffective proof-of-concept approach;
  - Ineffective communication process with vendor technical-support departments.
- Time gap between design and implementation phases (such as customer political and/or bureaucratic processes). This time gap leads us to the following issues:
  - Organizational problems: (1) changes in legislation (customs, tax, license and other rules) that prevent the equipment delivery within a reasonable time or make it impossible; (2) changes in a vendor pricing policy that make this equipment unattractive to the customer or lead to system architecture re-design (fixed system budget); (3) changes in vendor product lines.
  - Technical issues: (1) equipment problems – new bugs in new versions of software and/or hardware; (2) environmental problems – often some environmental components (cable, power supply or climate control systems) are not under our influence (we have to rely on results of related projects) and sometimes our expectations do not meet our needs (system requirements).

Most of these problems can be resolved by low-level re-design and, in bad cases, technical specification correction. However, in the worst case, some of these problems may have an impact on the high-level architecture design solution (up to total re-design).

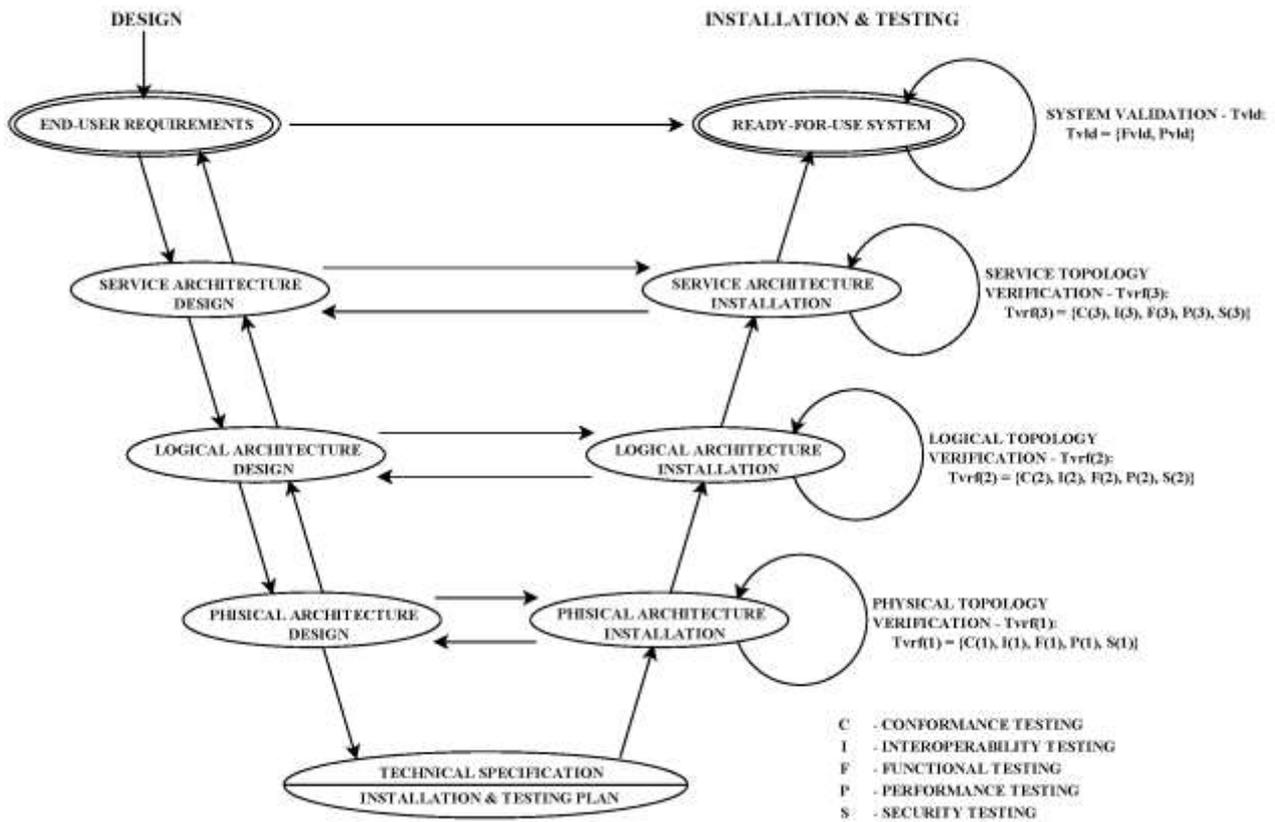

Fig. 1 The Adopted V-Model.

The resulting Adopted V-Model is shown in Fig. 1.

In the general case, the purpose of the systems verification testing process is to ensure that the system is ready for delivery [12] [15]. Theoretically verification testing has to include full coverage of the whole system:

$$T_{vrf} = \bigcup_{i=1}^{N} T_{vrf}(i)$$

where $T_{vrf}$ is a finite, nonempty set of system verification testing procedures; $N$ is the number of model hierarchical layers (in our case $N = 3$); and $T_{vrf}(i)$ is a finite, nonempty set of test cases on layer $i$. In turn:

$$T_{vrf}(i) = \{C(i), I(i), F(i), P(i), S(i)\}$$

where $C(i)$ is a finite set of conformance test cases on layer $i$; $I(i)$ is a finite set of interoperability test cases on layer $i$; $F(i)$ is a finite set of functional test cases on layer $i$; $P(i)$ is a finite set of performance test cases on layer $i$; and $S(i)$ is a finite set of security test cases on layer $i$.

On the over hand, the purpose of system validation testing is to determine whether or not a system satisfies its acceptance criteria [12] [15]. Validation testing does not include full coverage of the whole system. The intent is to demonstrate that critical, high-risk and complex capabilities of the system are working properly. Thus, validation testing processes are only a subset of the verification testing process:

$$T_{vld} \subseteq T_{vrf}$$

where $T_{vld}$ is a finite, nonempty set of system validation testing procedures.

Recommended limit of the set of test cases for the system validation testing [9]:
- from two to five test cases covering functionality;
- two or three test cases covering performance.

In this case:

$$T_{vld} = \{F_{vld}, P_{vld}\}$$

And

$$F_{vld} \subset \bigcup_{i=1}^{N} F(i)$$

$$P_{vld} \subset \bigcup_{i=1}^{N} P(i)$$

where $F_{vld}$ is a finite, nonempty set of functional test cases; and $P_{vld}$ is a finite, nonempty set of performance test cases.

## V. CONCLUSIONS

Deployment of network/distributed systems sets high requirements for procedures, tools and approaches for complex testing of these systems. In this work we provided a survey of testing activities with regard to these systems based on standards and actual practice. On the basis of this survey, we identified the most relevant approaches and related standards dealing with both distribution (network) and software-based systems aspects covering traditional and SOA testing issues. We also discussed relevant performance characteristics and security testing approaches. These formal testing procedures cover both software-based and distribution (network) aspects, and are not contrary to these aspects.

Next, based on the analysis of the implementation phase of SDLC, we determined a formal model of testing activities. The model allows the explicit stating of transitions of some activities that are often neglected but must be considered when more advanced automation processes are planned. The Adopted V-Model (see Fig. 1) defines the position of testing processes during Design/Implementation phases of SDLC and their interactions with other processes – design and installation. This model is adopted for the challenges of actual practice. Therefore, we listed a number of events that often occur in practice and cause additional dependencies and redirections of process flows.

As a consequence, we have identified a top-level framework covering all necessary and required activities dealing with complex testing of network/distributed systems.

ACKNOWLEDGMENT